\documentclass[twocolumn,
aps,nofootinbib,showpacs,showkeys,preprint
tightenlines
] {revtex4-1}

\usepackage{epsf,epsfig,subfigure,graphicx,amsmath,amssymb}
\usepackage{color}
\usepackage{float}
\newcommand{\dis}[1]{\begin{equation}\begin{split}#1\end{split}\end{equation}}

\newcommand{\gev}{\,\textrm{GeV}}

\newcommand{\etal}{{\it et al.}}

\newcommand{\EE}{{{\rm E}_8}}

\newcommand{\chSUsix}{{\rm SU(6)_{ch}}}
\newcommand{\SUfa}{{\rm SU(3)_{family}}}

\newcommand{\one}{{\bf 1}}
\newcommand{\oneb}{\overline{1}}
\newcommand{\ten}{{\bf 10}}

\newcommand{\five}{{\bf 5}}
\newcommand{\fiveb}{\overline{\bf 5}}
\newcommand{\fif}{{\bf 15}}

\newcommand{\six}{{\bf 6}}
\newcommand{\sixb}{\overline{\bf 6}}

\newcommand{\three}{{\bf 3}}
\newcommand{\threeb}{\overline{\bf 3}}
\newcommand{\two}{{\bf 2}}

\begin{document}


\title{\Large\bf  Supersymmetric three family chiral SU(6) grand unification model from
   F-theory}

\author{Kang-Sin Choi$^{a}$ and Jihn E. Kim$^b$
\email{jekim@ctp.snu.ac.kr}
}
\affiliation{
$^{a}$Korea Institute for Advanced Study, Seoul 130-722, Korea\\
$^b$
Department of Physics and Astronomy and Center for Theoretical Physics, Seoul National University, Seoul 151-747, Korea\\
 }
\begin{abstract}
We obtain a supersymmetric three family chiral SU(6) grand unification model with the global family symmetry $\SUfa$ from the use of F-theory.
This model has nice features such as all the fermion masses are reasonably generated and results in only one pair of Higgs doublets, giving the doublet-triplet splitting from the family symmetry  $\SUfa$. The proton hexality is also realized toward the proton stability problem.  The model also has a room to fit the three gauge couplings using the F-theory flux idea and we obtain the upper bound of proton lifetime in the $10^{36-37}$ yr region.
\end{abstract}


\maketitle

The standard model (SM) has two outstanding theoretical problems: the flavor problem and the mass problem of the Higgs boson. The flavor problem is a question; ``Why are there three repetitions of fermions and how do we distinguish them?" The scalar mass problem is also question, ``Why is the mass parameter of the Higgs boson at the electroweak scale while the SM can be a valid theory up to the Planck mass scale $M_P$?" To understand these problems, the SM symmetry has to be a part of a bigger symmetry that will allow the additional symmetry structure to dictate the number of families and the structure of Yukawa couplings, leading to a tiny Higgs mass parameter.

Since the SM fermion masses are below TeV, about $10^{-16}$ times $M_P$, the quark and lepton structure under the SM gauge group SU(3)$\times$SU(2)$\times$U(1) must be chiral. Therefore, it is better to have any extension of the SM being chiral from the beginning, as formulated by Georgi \cite{Georgi79}.

As compared to the Planck mass, the tiny Higgs boson mass requires the introduction of gravity in the SM extension. In this regard, the Higgs mass problem has been tried to be understood in the minimal supersymmetric (SUSY) extension of the SM, the so-called MSSM, formulated in the supergravity models.

However, in including both the above extensions at field theory level, one encounters a severe problem due to the proliferation of the needed matter fields. However, this dilemma has been resolved by introducing extra dimensions. With use of extra dimensions, the gravitational anomalies restrict possible gauge groups, in particular to $\EE\times \EE'$ and SO(32) in ten dimensions  \cite{GS84}. The heterotic string idea gives these groups and in string theory, we have a limit to the size of gauge symmetry to ten dimensions, primarily due to the gravitational anomalies which contributes a limit to the number of chiral fermions, presumably to three visible sector families. Therefore, string theory is a good candidate for use in the understanding of the flavor problem. Obtaining the three family MSSMs from string requires the compactification of six extra dimensions. Orbifold compactification has been extensively used toward obtaining the MSSM \cite{ChoiKSbook}.

In the MSSM, the strong, weak, and normalized weak-hypercharge gauge couplings seem to be unified around at $10^{16}$ GeV \cite{CoupUnif} with the bare value of the weak mixing angle at the grand unification (GUT) scale set at $\sin^2\theta_W^0=3/8$. As a  result, GUT models with $\sin^2\theta_W^0=3/8$ are appropriate \cite{Kim03}, though the heterotic string compactification generally lacks an adjoint scalar field in breaking the GUT down to the SM. This has led to the search of flipped SU(5) GUTs from heterotic string \cite{FlipSU5}.

On the other hand, F-theory gives another way of symmetry breaking in the adjoint direction without reducing the rank. This is done by turning on magnetic flux, whose field strength is inherited from Ramond--Ramond antisymmetric tensor by Kaluza--Klein reduction  \cite{Donagi09}.

A nice flavor unification in SU(11) \cite{Georgi79} contains an SU(6) GUT \cite{KimSU6} which must be chiral. In this paper, based on the F-theory framework, we construct Kim's three family $\chSUsix$ GUT model obtained from SU(11),
\dis{
\fif_L &=\left(\begin{array}{cccccc}
0&u^c& -u^c &u&d&D \\ -u^c&0&u^c & u&d&D \\
u^c&-u^c&0 &u&d&D \\ -u&-u&-u&0 & e^c&H_u^{+} \\
-d&-d&-d& -e^c&0 & H_u^{0} \\
-D&-D&-D&-H_u^{+}  & -H_u^{0} &0 \\
\end{array}\right),\\
\\
\sixb_L^{\,M} &=\left(\begin{array}{c}
  d^c\\ d^c\\ d^c \\  \nu_{e}  \\  -e \\  N
\end{array}\right),\
\sixb_L^{\,H} =\left(\begin{array}{c}
  D^{c}\\  D^{c}\\  D^{c}\\ H_d^{0}  \\  -H_d^{-} \\  N'
\end{array}\right).\label{eq:SUsixGUT}
}
An extra pair of Higgses $\sixb^{\,h}$ and $\six^{\,h}$ develop vacuum expectation values (VEVs) which are eaten by the superheavy gauge multiplet. The symmetry SU(6) is then broken down to SU(5), reproducing the standard SU(5) matter and Higgses.
We can see that any SU(5) interaction inherited from a renormalizable SU(6) invariant coupling preserves the $R$-parity.
The component having the 6th tensor index has the opposite $R$-parity with the rest, for example $\fif \to \ten^M + \five^H$, $\sixb^H \to \fiveb^H + \one^M$. This SU(5) is also broken by magnetic flux near the same energy scale to
the SM group. Since the electroweak hypercharge is ${\rm diag}.(\frac{-1}{3},\frac{-1}{3},\frac{-1}{3},\frac12,\frac12,0)$, the bare weak mixing angle ought to be $\frac38$ however, it is slightly corrected due to the flux as seen below.

This model is naturally obtained from F-theory compactified on elliptic Calabi--Yau fourfold  \cite{Vafa96}. Gauge symmetry is described by a singularity having the same structure of Lie algebra  \cite{Bershadsky96}. In the low-energy description, it is the gauge theory on eight-dimensional worldvolume theory \cite{Beasley09}.  Among them, four dimensions $S$ are inside Calabi--Yau manifold, supporting the SU(6) singularity while the other harbors our spacetime.
We embed instantons in SU(2) and SU(3) subgroups of $\EE$ in the internal space. Our unification group is the unbroken commutant SU(6).
The branching of the ${\bf 248}$ gaugino of E$_8$ into the representations of SU(6)$\otimes $SU(2)$\otimes $SU(3) gives
\dis{
 \text{adjoints}  \oplus [(\fif, \one,\three) \oplus (\sixb,\two,\three) +
 {\rm c.c.}]  \oplus ({\bf 20},\two,\one)\label{eq:EESplit}.
}
The ``off-diagonal'' components become matter fields, which are chiral as being zero modes of the background gauge bundle.
We denote the weights of $\bf 3$ of SU(3) as $t_1,t_2,t_3$ and $\bf 2$ of SU(2) as $s_1,s_2$. The $S_3$ monodromy shuffles $t_1,t_2$ and $t_3$ and we
split the SU(2) structure group into S[U(1)$\times$ U(1)].

In discussing particle phenomenology, it is more useful to discuss it in terms of the observable gauge groups. So, we look into the SU(6) subgroup of $\EE$.
The charge raising and lowering adjoint representations of SU(6), SU(2) and SU(3), as subgroups of E$_8$, are represented as (for instance, using Dynkin diagram strategy  \cite{Hwang03})
\dis{
&{\rm SU(6):}~(\underline{1~\oneb~0~0~0~0}~0~0); \\
&{\rm SU(2):}~T_\pm=(0~0~0~0~0~0~\underline{1~\oneb}); \\
&{\rm SU(3)_\perp:}~I_\pm=\pm(0~0~0~0~0~0~{1~1}),  \\
&\quad\quad\quad V_+= (-^6++),~ V_-=(+^6--),\\
&\quad\quad\quad U_+=(-^6--),~U_-=(+^6++),
}
where $\pm$ of spinors imply that $\pm\frac12$, and $\oneb=-1, \overline{2}=-2$, etc.

It is a remarkable fact that the same physics can be described in terms of geometry. The concrete background bundle, taking into account the SUSY and the splitting and monodromies, is provided by spectral cover \cite{Witten97}, described by
\dis{ \label{speccover}
 (b_2 s^3 + b_4 s + b_5)(d_0 s + d_1)(d_0 s - d_1)=0.
}
It is embedded in a projectivized fiber bundle $\check Z= \mathbb{P}(K_S \oplus {\cal O}) \stackrel{\pi}{\to} S$, where $K_S$ and $\cal O$ are the canonical and the trivial bundles on $S$, respectively. $S$ is located at $s=0$ where $s$ is an affine coordinate for the zero section $\sigma$. The coefficients are sections $b_m \sim (6- m)c_1-t-2x,d_m \sim (1-m)c_1+x$, where $c_1$ and $-t$ are respectively the tangent bundle of $S$ and the normal bundle to $S$ in the base of elliptic fibration. We have freedom to choose an integral two-cocycle $x$ on $S$ and a convenient choice is $x=-c_1$.

The matter fields are localized along distinct intersections between spectral covers and $S$, the so-called matter curves. Employing the standard procedure of finding irreducible common intersections between factors in Eq. (\ref{speccover}) \cite{ChoiKS10}, we can obtain the curves given in Table \ref{t:MatQN}

\begin{table}
\begin{tabular}{lll}
\hline
 matter & weight &  homology class of $\Sigma_R$ in $\check Z$  \\
\hline
 $\fif$ & $\{t_1,t_2,t_3\}$ & $(\eta-5c_1-2x)\cap \sigma$ \\
 $\sixb^{\,M}$ & $\{t_1+s_1,t_2+s_1,t_3+s_1\}$ & $(\eta-2c_1+x)\cap(\sigma+c_1+x)$\\
 $\sixb^{\,H}$ & $\{t_1+s_2,t_2+s_2,t_3+s_2\}$ & $(\eta-2c_1+x)\cap(\sigma+c_1+x)$\\
  ${\bf 20}$ & $\{s_1\}$ & $(c_1+x) \cap \sigma $\\
 ${\bf 20}^{\,\prime}$ & $\{s_2\}$ & $(c_1+x) \cap \sigma $ \\
 \hline
  \end{tabular}
\caption{Matter, defining curves and their homologies. We omitted the pullback.} \label{t:MatQN}
\end{table}

To have four dimensional chiral spectrum, we turn on the so-called $G$-flux, a magnetic flux along the Cartan subalgebra. Turning on the universal flux $\Gamma$ on the SU(3) spectral cover,
\dis{
\Gamma = \lambda(3\sigma-\eta+5c_1+2x),
}
the magnetic flux is induced on the matter curve.
The number of zero modes of Dirac operator is given by the Riemann--Roch--Hirzebruch index theorem \cite{Beasley09}
\dis{
 n_{R} - n_{\,\overline R} = \Sigma_R \cap \Gamma|_S = - \lambda \eta \cdot (\eta-3c_1)\label{eq:numberfam}
}
for ${\bf 15}, {\bf \overline 6}^{\,M},{\bf \overline 6}^{\,H}$, where the dot product is done on $S$. Since we turned on the flux in the SU(3) part only, the SU(2) charged components are blind to the flux, so $\sixb^{\,}$'s and $\fif$'s have the same number of zero modes. In particular, ${\bf 20}$'s are neutral under SU(3); thus they do not exist in four dimension. Choosing the base space for the discriminant locus such that (\ref{eq:numberfam}) is 3 for each matter representation, we can obtain three families (See, for example, \cite{ChoiKS10}).
For $\sixb^H$ we assume $n_{\sixb^H}=4$ and $n_{\six^H}=1$, for there is no known way of calculating the individual numbers although the cohomology is easily classified. This vectorlike pair is responsible for the breaking SU(6) down to SU(5).

We also turn on a line bundle $L$ in the direction $\text{diag.}(\frac{-1}{3},\frac{-1}{3},\frac{-1}{3},\frac13,\frac13,\frac13)$, breaking SU(6) to  SU(3)$_c \, \otimes$ SU(3)$_W \, \otimes$ U(1)$_6$. Its common intersection with the previous SU(5) is the SM group. Employing the mechanism used in the SU(5) GUT from F-theory \cite{Beasley09,Marsano09}, this flux removes colored triplet $D^c$ inside $\sixb^{\,H}$. This line bundle also corrects the number of generations for some of the SM fermions to (\ref{eq:numberfam}) by $\Sigma \cap \Gamma_L|_S$, where $\Gamma_L$ is the contribution from $L$ \cite{Blumenhagen:2009yv,Marsano09}; then, for three generations, we need to turn on fluxes on the other spectral covers to adjust the zero modes. In general it gives rise to an extra $\bf 20-\bf 20'$ pair, but they form a vectorlike mass and decouple. We need to take into account D-term contribution from the flux also.

If the MSSM leads to an exact unification, the flux is not allowed simply because it is against the data. We have about 4\% discrepancy in $\alpha_{3c}$ at the two-loop level, so we hope that an additional contribution from the flux can compensate for it. It was found to be not possible \cite{Blum09} provided the flux scale $f$ is {\em lower} than naive unification scale, since the relative strengths of the corrected coupling get a undesirable relation $\alpha_{3c} > \alpha_1 > \alpha_{2W}$,
where $\alpha_{3c},\,\frac35\alpha_1=\alpha_Y$ and $\alpha_{2W}$ are the fine structure constants respectively for the SU(3), SU(2), and U(1) groups of the SM.

However, we find the relation can become desirable in case where $f$ is {\em larger}. The flux corrects the couplings $\alpha_i^{-1}$ schematically as \cite{Blum09}
\dis{
 C_i^{-2} \alpha_i^{-1} = \alpha_{GUT}^{-1} + \alpha_{FW}^{-1} + D_i \alpha_{L}^{-1} ,
\label{eq:fluxcontrib}
}
where $\alpha_{GUT}$ is the field theoretic unified coupling without considering the flux, $\alpha^{-1}_{FW}$ is the corrections from the fluxes for global anomaly condition and $\alpha_{L}^{-1}$ is from $L$.
In our model, we also have $\alpha_{3W}$ of SU(3)$_W$  and $\alpha_6$
of the above U(1)$_6$.
We use the normalization $C_i^2$ with respect to the fundamental representation of SU(6), of couplings $\alpha_i$:
 $C_6^2=\frac43,\,C_3^2=1,\,C_{SU(3)_W}^2=1.$ Also $D_i$ parameterize the contribution from $L$, namely $D_6=\frac34,\, D_3=0,\, D_{SU(3)_W}=1$.
Although these corrections depend on moduli of internal space, we can eliminate such dependence by relating three couplings.

\begin{figure}[!t]
  \begin{center}
  \begin{tabular}{c}
   \includegraphics[width=0.35\textwidth]{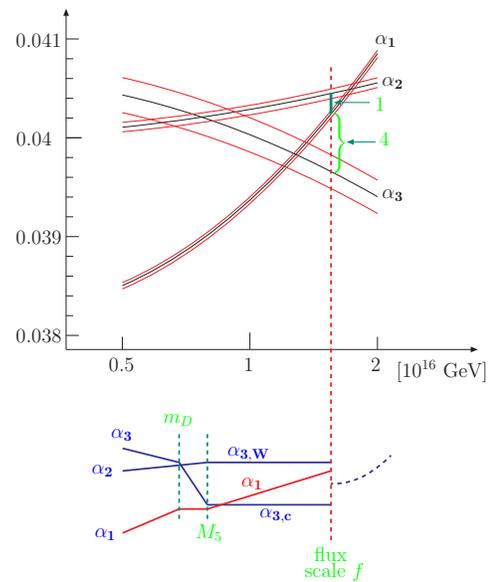}
   \end{tabular}
  \end{center}
 \caption{ Two loop evolution \cite{PData10} of couplings. Complete multiplets removed between $f$ and $M_5$ do not change the difference of couplings at $f$.
  }
\label{fig:DataDetail}
\end{figure}

The actual running above the triplet mass $m_D$ and Higgsing scale $M_5$ is drawn in the lower part of Fig. \ref{fig:DataDetail}.
Although we do not know these parameters, we can estimate $f$ as follows. In the limit where $m_D,M_6,f$ are coincident, we can use the field theoretical running of the gauge couplings from the electroweak scale known from the experiments \cite{PData10}, viz. the upper part of Fig. \ref{fig:DataDetail}. This sets the upper bound on $f$. In what follows, the gauge couplings mean the values at $f$.
Comparing the gauge kinetic functions of hypercharge and of SU(6), the gauge couplings satisfiy the relation
$ \alpha_1^{-1}  = \alpha_6^{-1} + \frac13 \alpha_2^{-1}$,
where we used $\alpha_2^{-1} = \alpha_{3W}^{-1}$ and the normalization for the SU(3)$_W$ breaking generator ${\rm diag}.(0,0,0,\frac16,\frac16,\frac{-1}{3})$.
Plugging them in (\ref{eq:fluxcontrib}), our model gives
\dis{
 ( \alpha_1-\alpha_3) : (\alpha_2 -\alpha_3) = 4:5.
}
From this and taking the universal SUSY scale as 1 TeV,
we fit the flux scale $f$ to the data from Fig. \ref{fig:DataDetail},
\dis{
f < 1.55\times 10^{16}\,\gev,
}
So the {\em upper bound} of proton lifetime is estimated as $(1-25)\times 10^{36}$ years \cite{LangGUT}. The SU(5) separation scale, $M_5\simeq m_D$, can be dialed to fit other aspects of the data.

We observe that there exist the quark and lepton Yukawa couplings needed for their masses, guided by the full $\EE$ gauge invariance.
The $Q_{\rm em}=\frac23$  quark masses arise from
$\fif \, \fif \, \fif$, recalling the definition (\ref{eq:SUsixGUT}).
The $Q_{\rm em}=-\frac13$ quark and charged lepton masses arise from
$\fif~ \sixb^{\,M}\sixb^{\,H}$. It also contains the Dirac neutrino mass by identifying the right-handed neutrino as the SU(5) singlet component $N'$ in $\sixb^{\,H}$.
Since $\nu$ and $N$ belong to the same multiplet $\sixb^M$, we do not have the Dirac mass between them.

The minimal invariant coupling containing the Majorana mass is
\dis{
 \frac{1}{M_P} \sixb^{\,H} \, \sixb^{\,H} \, \six^h \, \six^h.
\label{eq:Sixbdef}
}
The SU(6) breaking Higgs $\sixb^{\,h}$ (c.c. of $\six^h$ of Eq. (\ref{eq:Sixbdef})) should have the same quantum number as $\sixb^{\,H}$, hence the component $N'$ has odd $R$-parity.
Then the heavy neutrino Majorana mass is given by the following coupling
\dis{
\simeq \frac{V^2}{M_P}N'N'\label{eq:Vtilde}
}
where $\langle \six^h \rangle =V$.
Then, the $\nu N'$ seesaw mass matrix \cite{Minkowski77} has the conventional form
\dis{
\left( \begin{array}{cc}
0 & v_u \\
v_u & M_{N'}
\end{array}
\right)\label{eq:NuMassMatrix}
}
where $M_{N'} \sim { V^2}/{M_P}$. Then, the SM neutrino Majorana mass is obtained as
\dis{
m_\nu\simeq \frac{v_u^2}{M_{N'}}.
}
For $M_{N'}\approx 10^{14}$ GeV, we obtain $m_\nu\simeq 0.1$ eV which falls in the right ball park.

Finally, we comment on two important issues in the MSSM.
\vskip 0.1cm

{\it Proton hexality and proton lifetime}:
The GUT scale gauge bosons lead to proton lifetime at the level of $\sim \alpha_{\rm GUT}^{-2}\,f^4/{\rm GeV}^5\sim 10^{36-37}$ years.

But SUSY models with B and L violating $D=4$ operators lead to a disastrously short proton lifetime. Usually, it is forbidden by an $R$-parity or some kind of matter parity \cite{Ibanez91}. The most severe potential $D=4$ proton decay operator $u^c d^c d^c$ is contained in  $\fif ~ \sixb^{\,M} \sixb^{\,M}$.
It is not invariant under S[U(1) $\otimes$ U(1)] since $(t_1)+(t_2+s_1)+(t_3+s_1)=2s_1$ is nonvanishing. We might imagine an SU(6) invariant combination compensating $2s_1$ which develops large VEVs and induces a proton decay operator. The minimal one would be $\six^M \sixb^{\,h}$, but it does not induce a proton decay operator since the SU(5) singlet part of $\six^M$ does not develop a VEV. The field $\fif$ does not contain a neutral component under the SM group, and $\bf 20$ cannot form a gauge invariant operator leading to proton decay.

However, the exact $R$-parity allows $D=5$ superpotential term of the form $qqq\ell$ which are known to be dangerous for $M_{\rm GUT}\sim 10^{16}$ GeV.
This operator originates from the coupling $\fif \, \fif \, \fif ~ \sixb^M$. However, this is already forbidden by both SU(6) and SU(3) $\otimes$ S[U(1) $\otimes$ U(1)] invariance. We cannot use simple compensating combinations such as $\six^M$ or $\sixb^h \fif$, for the same reason as above. The same is valid for operators involving ${\bf 20}$. Antisymmetrizing $\sixb$'s always involves the SM charged fields.
Thus, we effectively obtained the proton hexality \cite{Dreiner06} in GUTs (see also \cite{Forste10}). One should admit that, although we can {\em describe} the vacuum yielding SM, its dynamical explanation is not possible before knowing detailed information about higher order potentials and moduli stabilization mechanism.

\vskip 0.2cm
{\it The $\mu$ problem and one pair of Higgs doublets}:
The Higgs doublets $H_u$ and $H_d$ form a vectorlike pair. Beyond the most important part of the $\mu$ problem, ``Why do/does the Higgs doublet pairs survive down to the electroweak scale?" \cite{KimNilles}, in the MSSM the next equally important problem is, ``Why is the surviving number of Higgs doublet pairs just one?" This second problem is suggested to be resolved using a bosonic family symmetry $\SUfa$ in Ref.
\cite{KimGMSBst}. It is similar to introducing {\em color gauge degrees} from the fermionic constituents ($u,d,s$) of baryons from the observation of the completely symmetric hadron wave function {\bf 56} in the flavor-spin space SU(6) \cite{HanNambu}. Our idea is based on the assumptions on three families. For three families, there are nine SU(3)$_W$ triplets from the quarks sector and there must be nine anti-triplets from the lepton and Higgs sectors. Out of nine SU(2) doublets resulting in the lepton and Higgs sectors below the GUT scale, six doublets are three vector-like pairs of $H_u$ and $H_d$ types. So, these Higgs doublets carry a family index $I=1,2,3$ in the SU(3)$_W$ basis. Since there is no $\three_I$, the family indices allow the superpotential of the form $\epsilon_{IJK}\threeb^{\,I} \threeb^{\,J}\threeb^{\,K}$. This leads to an antisymmetric $3\times 3$ matrix for the Higgsino masses with one mass being zero at the GUT scale. It is an elegant resolution toward the {\it one pair problem}. This has a seed in $\SUfa$ symmetry as realized in our F-theory.

This leads to the question, ``Then, how does the mass hierarchy of quarks and leptons arise?" The coupling, $\fif \, \fif \, \fif$,
contracts $\SUfa$ indices in the way $u^{cI} u_I$ and there is no $\epsilon_{IJK}$ symbol. Also, $\fif~\sixb^{\, M}\sixb^{\,H}$ does not involve $\epsilon_{IJK}$. As a result, the quark and lepton mass matrices are not anti-symmetric, and a reasonable mass hierarchy of quarks and leptons is obtained.

\acknowledgments{We thank Do-Young Mo and Seodong Shin for checking the running of gauge couplings. This work is supported in part by the National Research Foundation  (NRF) grant funded by the Korean Government (MEST) (No. 2005-0093841).}

\vskip 0.5cm

\vskip 0.5cm

\end{document}